\documentclass[12pt,preprint]{aastex}
\usepackage{natbib}
\usepackage{amssymb}

\newcommand{\bdv}[1]{\mbox{\boldmath$#1$}}

\newcommand{\cf}{{\it cf.\ }}
\newcommand{\mas}{{\rm mas}}
\newcommand{\kpc}{{\rm kpc}}

\def\bmu{{\bdv \mu}}
\def\rel{{\rm rel}}
\def\masyr{{\rm mas\:yr^{-1}}}
\def\e{{\rm E}}
\def\hatn{{\bf\hat n}}
\def\/{\_\nolinebreak[0]}

\begin{document}
\title{New Method to Measure Proper Motions of Microlensed Sources:
Application to Candidate Free-Floating-Planet Event
MOA-2011-BLG-262\footnote{Based on observations obtained with the 1.3-m Warsaw telescope at the Las Campanas Observatory of the Carnegie Institution for Science.}}

\author{
Jan Skowron\altaffilmark{1,2}, 
Andrzej Udalski\altaffilmark{1},
Micha{\l} K. Szyma{\'n}ski\altaffilmark{1},
Marcin Kubiak\altaffilmark{1},
Grzegorz Pietrzy{\'n}ski\altaffilmark{1,3},
Igor Soszy{\'n}ski\altaffilmark{1},
Rados{\l}aw Poleski\altaffilmark{1,2},
Krzysztof Ulaczyk\altaffilmark{1},
Pawe{\l} Pietrukowicz\altaffilmark{1},
Szymon Koz{\l}owski\altaffilmark{1},
{\L}ukasz Wyrzykowski\altaffilmark{1},
and
Andrew Gould\altaffilmark{2}
}
\email{(jskowron, udalski, msz, mk, pietrzyn, soszynsk, rpoleski, kulaczyk, pietruk, simkoz, wyrzykow)@astrouw.edu.pl and gould@astronomy.ohio-state.edu}
\altaffiltext{1}{Warsaw University Observatory, Al. Ujazdowskie 4,
                 00-478 Warszawa, Poland}
\altaffiltext{2}{Department of Astronomy, Ohio State University,
                 140 W. 18th Ave., Columbus, OH 43210, USA}
\altaffiltext{3}{Universidad de Concepci{\'o}n, Departamento de Astronomia,
                 Casilla 160--C, Concepci{\'o}n, Chile}

%%%%%%%%%%%%%%%%%%%%%%%%%%%%%%%%%%%%%%%%%%%%%%%%%%%%%%%%%%%%%%%%%%%%%%%%%
%                               Abstract
%%%%%%%%%%%%%%%%%%%%%%%%%%%%%%%%%%%%%%%%%%%%%%%%%%%%%%%%%%%%%%%%%%%%%%%%%
\begin{abstract}

We develop a new method to measure source proper motions in microlensing
events, which can partially overcome problems due to blending. 
It takes advantage of the fact that the source position
is known precisely from the microlensing event itself.  
We apply this method to the event MOA-2011-BLG-262, which has a
short timescale $t_\e=3.8\,$day, a companion mass ratio
$q=4.7\times 10^{-3}$ and a very high or high lens-source relative
proper motion $\mu_\rel=20\,\masyr$ or $12\,\masyr$
(for two possible models).  These three characteristics imply that the
lens could be a brown dwarf or a massive planet with a roughly
Earth-mass ``moon''.  The probability of such an interpretation
would be greatly increased if it could be shown that the high
lens-source relative proper motion was primarily due to the lens
rather than the source.  Based on the long-term monitoring data of
the Galactic bulge from the Optical Gravitational Lensing Experiment (OGLE), 
we measure the source proper motion that is
small, $\bmu_s = (-2.3, -0.9)\pm (2.8,2.6)\,\masyr$ in a (North, East) Galactic
coordinate frame.  These values are then important input into a Bayesian
analysis of the event presented in a companion paper by Bennett et al.

\end{abstract}


%%%%%%%%%%%%%%%%%%%%%%%%%%%%%%%%%%%%%%%%%%%%%%%%%%%%%%%%%%%%%%%%%%%%%%%%%
%                           Introduction
%%%%%%%%%%%%%%%%%%%%%%%%%%%%%%%%%%%%%%%%%%%%%%%%%%%%%%%%%%%%%%%%%%%%%%%%%
\section{{Introduction}
\label{sec:intro}}

Lens-source relative proper motions $\mu_\rel=|\bmu_l-\bmu_s|$
are frequently measured in planetary microlensing events,
but to date there are no published measurements of the source
proper motion itself in these events ($\bmu_s$).  This may seem surprising
at first sight because the source is almost always visible whereas
the lens is typically invisible.  In fact, however, $\mu_\rel$ is
generally both more useful and easier to measure than $\bmu_s$.

Source-lens proper motions can be measured essentially whenever there are
significant finite-source effects in the event \citep{gould94}
because the source-lens crossing time $t_*$ is directly measurable
from the light curve, while the angular size of the source can
be extracted from its dereddened color and magnitude \citep{albrow99},
which in turn can be extracted by placing the source on an instrumental
color-magnitude diagram \citep{yoo04}.  The most important application
of $\mu_\rel$ is not the proper-motion itself, but rather that it
immediately yields the Einstein radius,
\begin{equation}
\theta_\e = \mu_\rel t_\e = \sqrt{\kappa M \pi_\rel};
\qquad \kappa\equiv {4 G\over c^2\,\rm AU}
\simeq 8.1\,{{\rm mas}\over M_\odot},
\label{eqn:thetae}
\end{equation}
where $t_\e$ is the Einstein timescale (measurable from the event),
$M$ is the lens mass, and $\pi_\rel=\pi_l - \pi_s$ is the lens-source
relative parallax.  Therefore, $\theta_\e$ usefully constrains a
combination of the lens mass and distance.

However, $\mu_\rel$ does often play a role at the next level.
Because  $M$ and $\pi_l$ are not determined independently, one
normally must make a Bayesian estimate of these quantities, using
inputs from a Galactic model \citep{alcock97}, which can include
priors on $\mu_\rel$.  In principle, the Bayesian analysis could also
include priors on $\bmu_s$ if this quantity were measured.  
There are two reasons why this has not been done yet.  First, in many cases,
the posterior probabilities would not be strongly impacted by this
additional prior.  Second, and probably more important, it is
remarkably difficult to measure $\bmu_s$ in most cases.

Here we present a new method to measure $\bmu_s$, which is
tailored to meet the challenges of the faint, moderately blended sources
typical of microlensing events seen toward the Galactic bulge.  We are
motivated to develop this method by, and specifically apply it to,
the planetary microlensing event 
MOA-2011-BLG-262\footnote{\url{https://it019909.massey.ac.nz/moa/alert/display.php?id=gb8-R-7-4659}}/OGLE-2011-BLG-0703\footnote{\url{http://ogle.astrouw.edu.pl/ogle4/ews/2011/blg-0703.html}}.
This event has a short timescale $t_\e=3.8\,$days, a very high (or high) 
relative proper motion $\mu_\rel=19.6\,\masyr$ (or $11.6\,\masyr$
for the competing, nearly equally likely, microlensing model), and 
a companion/host mass ratio $q=4.7\times 10^{-3}$ \citep{bennett14}.  
These parameters are, in themselves,
consistent with either a lens that contains a stellar host
with a Jovian-class planet in the Galactic bulge, or a brown dwarf
(or possibly a Jovian planet) with an Earth-class ``moon''.

In the former case (stellar host in the bulge), 
the very high $\mu_\rel$ that is measured in the microlensing event
would almost certainly require combinations of abnormally
high lens and source proper motions.  That is, if the source were
moving slowly, it would be quite unusual for the lens proper motion
to be large enough to account for the high relative proper motion
by itself.  By contrast, if the lens were in the foreground disk
(and so of much lower mass),
its proper motion relative to the frame of the Galactic bulge
could easily be high enough to explain the observed $\mu_\rel$.
Thus, for this event, it would be important to actually
measure $\bmu_s$.

The Optical Gravitational Lensing Experiment (OGLE) is
a long-term photometric sky survey
focused on finding and characterizing microlensing events in the
Galaxy. 
The first phase of 
the OGLE project began in April 1992 and the project continues to this
date with its fourth phase currently being executed 
\citep{udalski03}. 
OGLE monitors 
brightness of hundreds of millions of stars toward the Galactic 
bulge with high cadence, using dedicated 1.3m Warsaw Telescope 
at Las Campanas Observatory, Chile. 
Every night between
100 and 200  1.4 deg$^2$ exposures are being taken. In addition to
the performed real-time reduction and analysis, 
all the science frames are archived in their full integrity.
These constitute an unprecedented data set for various 
astronomical studies.

The decades-long time span of the OGLE monitoring
provides us with a unique opportunity to precisely measure proper 
motions of many stars in the Galactic bulge, including
the source star of this very interesting microlensing event.

Proper motion studies of the
Galactic bulge have been previously carried out using the OGLE data.
For example, \citet{sumi04} measured proper motions for over $5\times 
10^6$
stars over 11 deg$^2$ using the OGLE-II data from 1997 to 2000.  However,
this survey was restricted to resolved stars, $I<18$.  

In the
present case, the source magnitude is $I\approx 19.9$ (as determined from the
microlens models), which would be close to the 
photometric detection limit even if
the source was relatively isolated.  In fact, the source is blended
with a brighter star, and was not recognized as an independent star
in the reference image, prior to the event.  Hence, a new technique is
required to measure the proper motion, which is described in
the next section.
%Section~\ref{sec:method}.  

\section{{New Method For Measuring Proper Motions}
\label{sec:method}}

\begin{figure}
\plotone{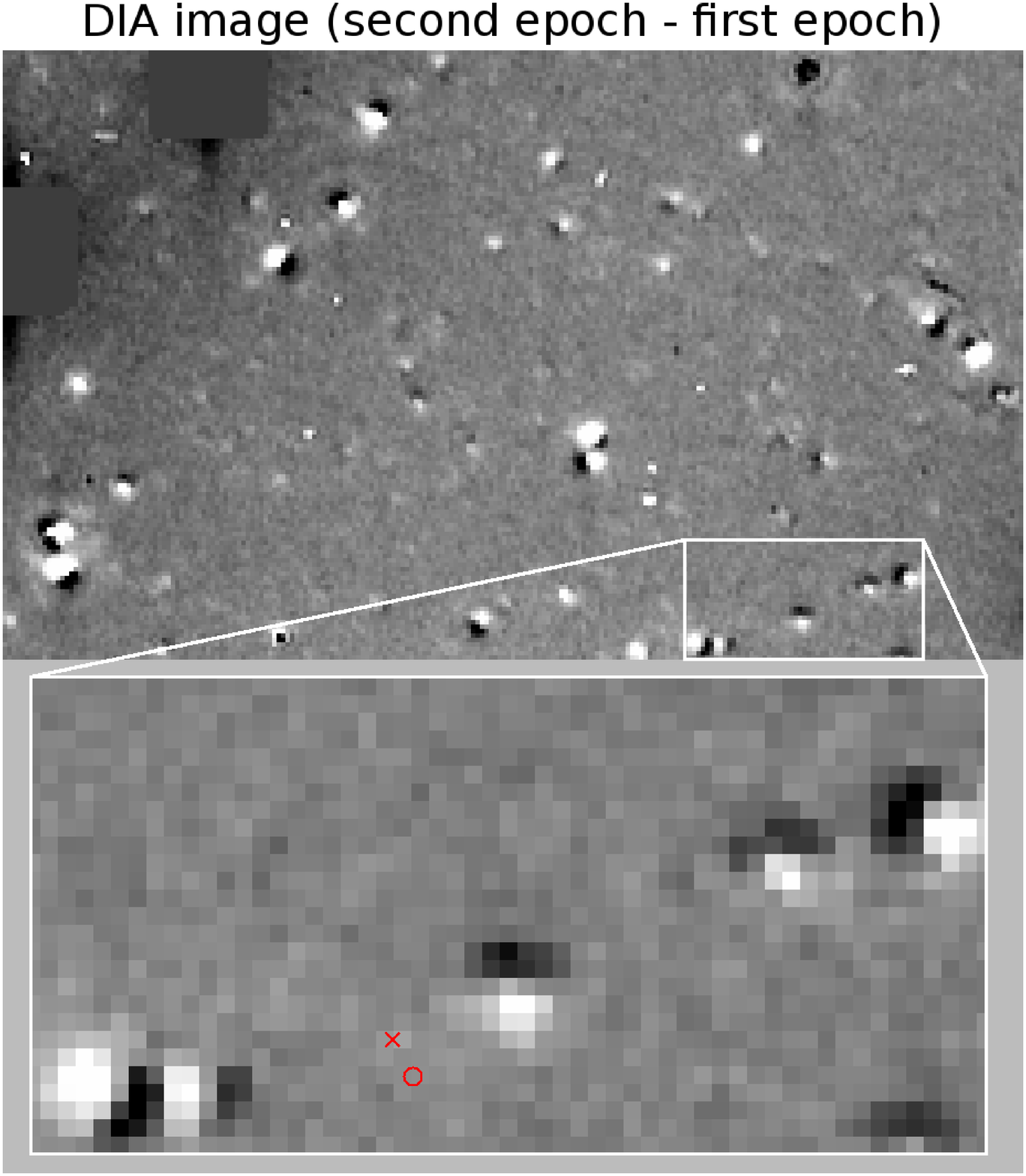}
\figcaption{The DIA image of the the first epoch
image convolved and subtracted from the second epoch image.
Two panels show $1'\times 0.6'$ and $14''\times 7''$ region.
The dipole-shaped profiles are visible throughout the
image.
Position of the microlensing event is marked with ``x''. Unrelated
neighboring star is marked with ``o''. 
North is up and East is to the left.
Difference between black and white colors is 130 counts. 
Background noise is 9 counts.
\label{fig:dia}
}
\end{figure}

Consider a difference of two images that has been generated by
standard difference image analysis (DIA,\citealt{woz00}).  
That is, the images have been geometrically
aligned to a common frame of reference stars, photometrically aligned
to the same mean flux level, and one has been convolved with a kernel
function
to mimic the point spread function (PSF) of the other.  The usual
purpose of this procedure is to detect stars whose flux has changed
between the two epochs.  These will appear as relatively isolated
PSFs on an otherwise flat background (beside the noise, cosmic rays,
masked regions, satellites, etc.).
However, let us now consider the
case that there have been no flux changes but only position changes.
For simplicity, we begin by assuming that the PSF is an isotropic
Gaussian
\begin{equation}
P(x,y) = {\exp(-r^2/2\sigma^2)\over 2\pi \sigma^2};
\qquad r^2=x^2+y^2.
\label{eqn:psf}
\end{equation}
where $\sigma = {\rm FWHM}/\sqrt{\ln 256}$ is the Gaussian width.
Let us now assume that one star has been displaced by a distance $\Delta r$
in the direction $\hatn$, while all other stars have remained fixed.
If we further assume that $\Delta r\ll\sigma$,
then it is straightforward to show that the difference profile will
have the form
\begin{equation}
D(x,y) = (f\Delta r)\hatn\cdot\nabla P
= (f\Delta r)(x\cos\phi + y\sin\phi){P(x,y)\over \sigma^2},
\label{eqn:dipole}
\end{equation}
where $\hatn = (\cos\phi,\sin\phi)$ and $f$ is the flux of the star.
Thus, the difference image will have an anti-symmetric dipole profile, 
whose
form is always the same ($xP$).  The amplitude of this dipole is
given by the product $f\Delta r$, and the direction is simply the
direction of motion.  Note that the maximum of the dipole profile
lies at $\sigma\hatn$, i.e., $1\,\sigma$ from the star center.
Moreover its height relative to the peak of the original star is
\begin{equation}
{{\rm max}(D)\over {\rm max}(f P)} = e^{-1/2}{\Delta r\over\sigma}
\label{eqn:max}
\end{equation}
Thus, if several stars have moved, then the difference
image will contain several dipoles, all with the same form, but with
different amplitudes and pointing in different directions.
See Figure~\ref{fig:dia}.

In principle, then, it is possible to determine the proper motion of a
star by measuring the height of the dipole relative to the original
image and applying Equation~(\ref{eqn:max}).  In practice, the utility
of this approach is limited to fairly restricted conditions.  First,
if the star is truly isolated, it is usually more convenient
to simply measure the star at many epochs
in the usual fashion.  On the other hand, if the star is heavily blended,
it will still produce a dipole as given by Equation~(\ref{eqn:dipole})
but there will be two problems.  First, the star's flux $f$ cannot
usually be disentangled from that of other stars within the PSF.
Hence, when the dipole amplitude ($f\Delta r$) is measured, one cannot 
extract
$\Delta r$ from it.  Second, other stars within the PSF may also have
moved, each in its own direction $\hatn_i$ and each with own amplitude
$(f\Delta r)_i$.  When several stars are in the same PSF, all
that can be observed is the vector sum of these dipoles:
$\sum_i (f\Delta r)_i\hatn_i$.  

Now, for microlensed sources, the first of these problems is actually
easily solved because the microlensing fit returns the source flux
$f_s$ as one of its parameters.  However, the second problem remains.
This means that the technique is actually applicable only to moderately
blended sources in which the blending can be reasonably well understood.

Before undertaking such an application, we note that the above formalism
is easily extended to asymmetric Gaussian profiles,\
\begin{equation}
P({\bf x}) = {\sqrt{|b|}\over 2\pi}\exp(-x_i b_{ij} x_j/2),
\label{eqn:psf_asym}
\end{equation}
 where
$b_{ij}$ is the inverse covariance matrix, ${\bf x}$ is now the
2-dimensional coordinate, and where we use Einstein summation convention.
Then the dipole is given by 
\begin{equation}
D({\bf x}) = (f\Delta r)\hatn_i b_{ij} x_j P({\bf x})
\label{eqn:dipole_asym}
\end{equation}
Of course, one might also consider non-Gaussian profiles, but since
almost all the signal comes from within a few $\sigma$, while the
deviations from a Gaussian are not understood well enough
to characterize $\nabla P$ at sub-pixel scales, this level
of complexity is generally not warranted.

\section{{Absolute reference frame of the proper motion measurement}
\label{sec:frame}}

The difference image in DIA can be constructed either by taking
the second epoch image as an ``image'' and the first epoch image as a 
``reference image'' or vice versa. 
In the latter case, the measured dipole direction 
should be flipped to obtain the real direction of the proper motion. 
Since the ``reference image'' will be convolved with a
kernel to match its PSF to the PSF of the ``image'', one should choose 
the lower-seeing image out of the two for the reference image.

Before the DIA procedure can be performed, the images should be aligned to 
a common pixel grid. This is done by picking a set of bright common stars
in both images, finding polynomial or spline interpolation of
the coordinates and resampling one image to the other image's pixel grid
\citep{woz00}.
This step ensures that both images cover the same region of the sky
-- modulo the quality of the transformation.
The differences should be small compared to the fitting regions 
of the DIA kernel (domains).
Other advantage this procedure brings is compensation of the majority 
of the differences of field distortion between the two epochs. 
Thus, the DIA kernel can have less free parameters to achieve 
similar quality of subtraction.

It is extremely important to understand what degrees 
of freedom the DIA kernel has before any 
conclusions about the proper motion can be made. 
One example would be a kernel transforming a Gaussian PSF
into a Gaussian PSF with the same centroid.
If the images were not carefully aligned to remove all 
of the field of view deformations the subtraction could
turn out poor. By introduction of the additional parameters 
to the mathematical model of the PSF, describing 
smooth shifts of its center across the field, subtraction
would be greatly improved.

\citet{woz00} DIA uses a series of Gaussians multiplied 
by polynomials for the mathematical model of the PSF. 
The difference of two Gaussian profiles displaced by 
a small distance $x$ ($x \ll \sigma$) is equal to 
the Gaussian multiplied by the first order polynomial
(\cf Eq.~(\ref{eqn:dipole})). Hence, if we allow a PSF model
to be constructed from a Gaussian multiplied by 
the first order polynomial, with parameters
being a function of position on the image, we effectively 
allow for the smooth profile centers displacements
(by some non negligible
fraction of the profile width).
This approach allows for compensation of the field distortions,
but at the same time, it potentially smoothes the gradients 
in the stars' motions across the field. 

Let us consider a globular cluster covering some 
fraction of the field in front of the Galactic background.
Stars in the cluster would have different expected median 
velocities than the background stars.
If, in a region of the frame, the percentage of
cluster members is significant, their motion between two epochs
could be seen as a field deformation near this region. 
Hence, careful initial alignment of 
the images should only use the field stars.
We also note, that the DIA is specifically designed to reduce 
signal on the subtracted image within the freedom of 
the PSF model.
Thus, without care in setting up the DIA procedure,
it is very likely, that the kernel 
parameters will try to compensate for the effective bulk 
motion of the part of the field containing cluster members 
-- effectively reducing observed dipole signal for the cluster
members and introducing false signal in the field stars.

In the field of view studied in this paper we do not expect 
any unusual velocity gradients. However, we expect that 
the bright stars sample would consist of a mixture of bulge stars
and disk stars. Disk stars are expected to have on
average proper motions of couple of $\masyr$ in the direction 
of Galactic rotation, with respect to the bulge stars.
For a reference point of our proper motion measurement
we choose the median proper motion of the Red Clump (RC) giants,
which are identified on the Color-Magnitude Diagram (CMD).
We use the RC giant stars for two reasons. 
First, they belong to the bulge system so their 
median proper motion is not affected by the Galactic rotation
and can serve as an approximation of the median motion of
the whole bulge system.
Second, the Red Clump giants are bright, abundant and easy to 
identify on the CMD.

We initially align two epoch images using the predefined sample 
of RC giants. 
After performing the DIA, we measure the shift of 
centroids of those stars that is likely  
introduced by the procedure. 
We take the ``reference'' epoch and measure 
the positions of the reference stars.
Then, we take ``convolved reference image'' 
(it is produced by the DIA in order to match 
the reference image to the target image 
before the subtraction)
and measure the 
positions of the same sample of stars.
The median shift between the measured positions 
indicates the total displacement the DIA procedure 
introduced, and hence, should be subtracted from any
dipole measurements to form the 
final proper motion measurement.

\section{{Application to MOA-2011-BLG-262/OGLE-2011-BLG-0703}
\label{sec:app}}

\begin{figure}
\plotone{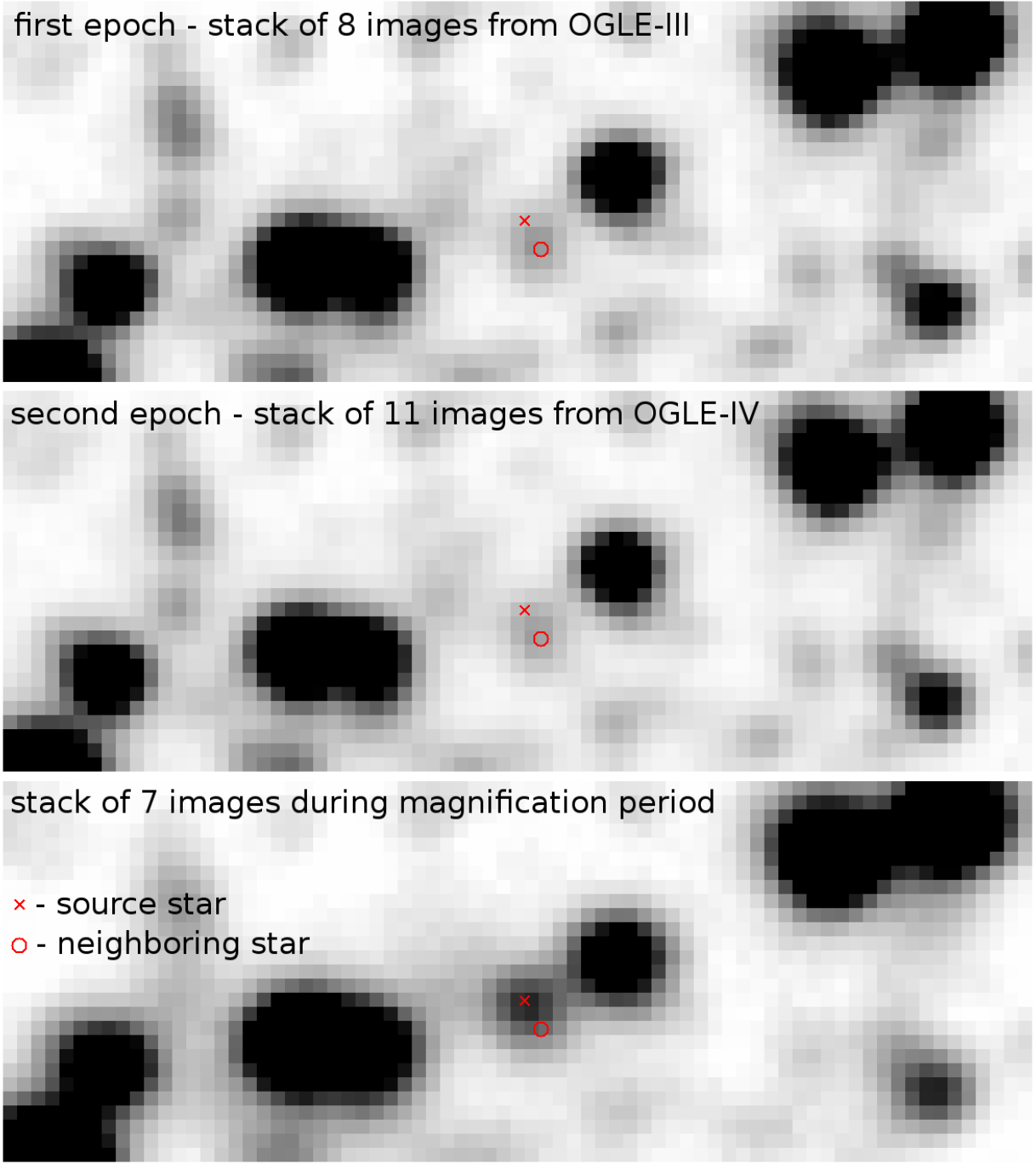}
\figcaption{Upper panel: image composed from 8 best-seeing frames from OGLE-III
(mean epoch 2669.0).
Middle panel: image composed from 11 best-seeing frames chosen from OGLE-IV
(mean epoch 5462.7).
Bottom panel: image composed from 7 observations taken during the microlensing
event when the source was magnified by $\sim 8$ ($HJD' \approx 5739.5$). 
Position of the microlensing event is marked with ``x''. Unrelated 
neighboring star, which is twice as bright as the source star, 
is marked with ``o''. Images cover $19'' \times 7''$ region; pixel scale is 0.259 "/px.
North is up and East is to the left.
\label{fig:images}
}
\end{figure}

We use a series of images obtained with 
the Warsaw Telescope during the third and fourth phase
of the OGLE project. 
We select eight
best seeing images near the beginning of the OGLE-III
(mean epoch $HJD'=$ $HJD-2450000=$ $2668.98$) and 11 of the best seeing (unmagnified) images
from OGLE-IV (mean epoch $HJD'=$ $5462.72$).  
Note that OGLE-III and OGLE-IV
have identical pixel scales: 259 mas px$^{-1}$.
We stack each set of images.  These stacked images, which
have similar, but not identical seeing FWHM $\sim 3.5\,$px,
are offset by 7.65 years in mean epoch.  They are presented
in Figure \ref{fig:images}.

As shown in Figure \ref{fig:images}, the source of this event
is moderately blended with a brighter neighboring star even in
the excellent-seeing images used to construct the two stacks.
In fact, there is one image with extremely good seeing on which
the source appears isolated, but the single-to-noise ratio (S/N)
of a single image is too low to perform accurate astrometry or
photometry.  Instead, we find the microlensing source position from difference
images at the time of the magnification event (as is standard practice),
when it can be easily observed (\cf Fig~\ref{fig:images}).
Then, we determine the source flux ($f_s$) from the microlens fit and subtract
the resulting source profile from the second-epoch stacked image
in order to investigate the structure of the blending star(s) alone
\citep{gould02} -- namely, measure positions and brightness.
%  See Figure~\ref{fig:images}.

There is no detectable remaining
flux at the source position.  We thereby place an
upper limit on the blend flux at the source position 
of $f_b<0.3 f_s$, implying
$I_b>21.1$. For stars in the bulge, this implies absolute magnitude to 
be $M_{I,b}>5.0$ (where we take the reddening $A_I=1.6$).
In particular, this allows only for bulge lenses with $M<0.9\,M_\odot$.
Using Equation~(\ref{eqn:thetae}), the measured $\theta_\e= 0.23\,\mas$ 
($0.14\,\mas$) for the very-high $\mu_\rel$ (high $\mu_\rel$) model
implies that the allowed mass range corresponds to lens-source separations
$D_{ls}>0.40\,\kpc$ ($D_{ls}>0.17\,\kpc$).  In the case of the former
(very-high $\mu_\rel$) model
the substantial fraction of the available
phase space is ruled out, thus putting significant constraints upon 
(but certainly not ruling out) a bulge lens.  In addition, this flux limit restricts
the presence of distant hosts in ``free-floating-planet'' scenarios.
For example, if the primary lens is assumed to be a 10-Jupiter-mass object,
then it would have a relative parallax 
$\pi_\rel =\theta_e^2/\kappa M = 0.5\,$mas, and therefore be at a distance
$D_l\sim 1.6\,$kpc.  The flux limit would then imply $M_I>8.6$,
i.e., $M_{\rm distant\,host}<0.35\,M_\odot$ for the putative host.

\begin{figure}
\plotone{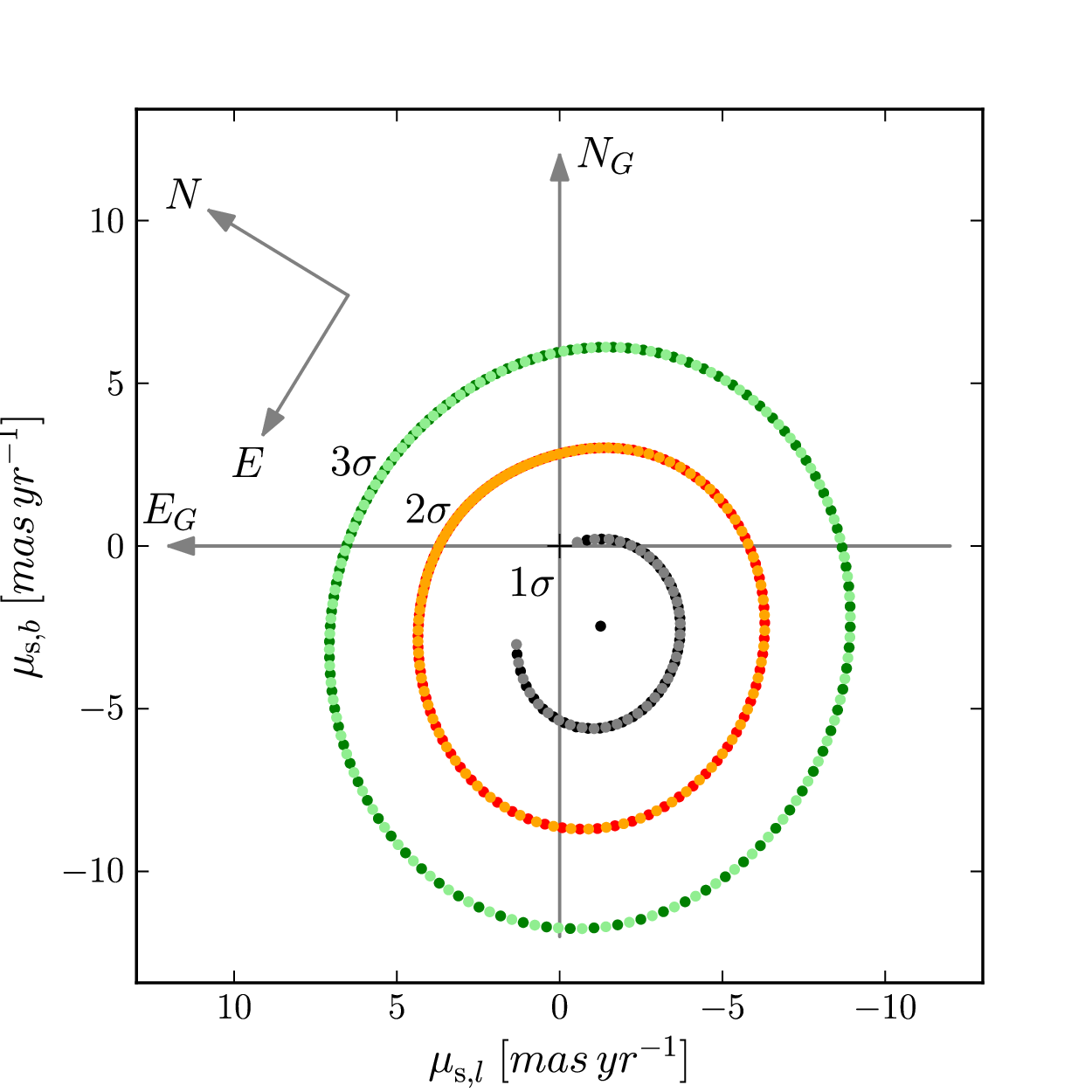}  % epoch 2-1 
\figcaption{Proper motion measurement of the source star
in MOA-2011-BLG-262 microlensing event ($\bmu_s$)
based on the DIA image, where the first epoch image (OGLE-III) 
was subtracted from the second epoch image (OGLE-IV).
Axes show direction of North Galactic pole ($N_{\rm G}$) 
and direction of Galactic rotation ($E_{\rm G}$).
Units of the plot are $\masyr$. 
Arrows in the upper-left corner show North and East direction
on the sky in the equatorial coordinates.
Contours mark one, two and three sigma regions, while black
dot shows the position of the $\chi^2$ minimum.
\label{fig:pm}
}
\end{figure}

However, our central concern here is to measure or place limits on the proper
motion of the source, $\bmu_s$.  We notice that there is no evident
dipole at the position of the source in the difference image in
Figure~\ref{fig:dia}.   To place limits, we must fit
for a dipole at this location.  Because the source is partially
blended with the neighboring star, we must fit simultaneously
for two dipoles, one at each location.  In fact, to be conservative,
for each (2-D) trial value of the source dipole, we consider all
possible values for the neighbor dipole, and choose the one that gives
the best overall fit.  We assign the resulting $\chi^2$ to this
trial value.  Each dipole amplitude $(f\Delta r)$ is divided by the
flux (known from the microlens fit) to obtain the displacement $(\Delta r)$,
and hence the proper motion $\bmu_s = \hatn\Delta r/ \Delta t$,
where $\Delta t=7.65\,$yr is the time elapsed between the two mean
epochs.  To model the PSF we use Equations~(\ref{eqn:psf_asym}) and
(\ref{eqn:dipole_asym}) with $b = (0.515,-0.008,-0.008,0.464)\,\rm px^{-2}$,
as measured with the DoPhot photometry package \citep{schechter93},
with the $x$-axis aligned to that of the OGLE-III camera, i.e., equatorial
North-South.  The resulting contours are shown in Figure~\ref{fig:pm}.

Origin in Figure~\ref{fig:pm} was adjusted for the shift
that was introduced by the DIA (see Section~\ref{sec:frame})
so it is now consistent with the median proper 
motion of the bugle stars.
In our $4.4' \times 4.4'$ work subfield we have identified 511 RC giants. 
The median shift of this set of bulge stars was measured 
between ``reference  images'' and ``convolved reference image''
to be
(0.0095, 0.0174) pixels in (North, East) direction. 
This is the equivalent to 
(-0.336,  0.583) $\masyr$ proper motion in the Galactic North
and East direction.

To test this shift-correction procedure we artificially resample 
target image by introducing simple shifts with values between 
$-0.1$ and $0.1$ pixels in both axes. The resulting subtracted image is virtually 
the same and yields same measurements of the proper motion
of the source star, showing that the DIA kernel easily absorbed
the artificial shift. The value of the displacement measured 
by the shift-correction procedure, recovers both, the original values 
quoted above, plus any artificial shift we have introduced.

We note that the contours shown in Figure~\ref{fig:pm} 
are elongated along an axis that is
approximately aligned with Galactic North-South.  This is because
the neighboring star lies along the Galactic East-West axis.
We can understand this analytically by approximating the PSF as
axisymmetric.  Without loss of generality, we can then assume that
the two stars are separated by a distance $a$ along the $x$-axis.  Their
dipoles can then be represented (in the global coordinate system) by
\begin{equation}
D_\pm = {[(x\pm a/2)\cos\phi_\pm + y\sin\phi_\pm]
\exp(-[-(x\pm a/2)^2 + y^2]/2\sigma^2)\over 2\pi\sigma^4}
\label{eqn:dip_pm}
\end{equation}
where $\phi_\pm$ are the orientations of the two dipoles relative to
the direction of their separation.
One then finds $\langle D_+|D_+\rangle = \langle D_-|D_-\rangle =
(4\pi\sigma^4)^{-1}$
and
\begin{equation}
{\langle D_+|D_-\rangle \over \langle D_+|D_+\rangle}
= -{[(a^2/2\sigma^2)-1]\cos\phi_+\cos\phi_- +\sin\phi_+\sin\phi_- 
\over \exp(a^2/4\sigma^2)}
\label{eqn:diracrat}
\end{equation}
It is straightforward to show that for uniform noise per pixel
(background-limited case) the ratio of the correlated errors
to the error estimates that would hold for an isolated source are
\begin{equation}
{\cal R}\equiv {\rm error (correlated)\over error (isolated)}
= \Biggl( 1 - {\langle D_+|D_-\rangle^2 \over \langle D_+|D_+\rangle^2}
\Biggr)^{-1/2}
\label{eqn:corerr}
\end{equation}

Then, for each possible orientation of the source dipole $\phi_+$,
we must choose the orientation of the neighbor-dipole that maximizes
Equation~(\ref{eqn:diracrat}) [and so Eq.~(\ref{eqn:corerr})].
Differentiation yields, $\cot\phi_- = [(a^2/2\sigma^2)-1]\cot\phi_+$,
from which one obtains,
\begin{equation}
{\cal R}(\phi_+) 
= \biggl\{1 - [1 - (1-k^2)\cos^2\phi_+]e^{-(1+k)}\biggr\}^{-1/2},
\qquad
k\equiv {a^2\over 2\sigma^2} - 1 .
\label{eqn:errcont}
\end{equation}

It is clear from Equation~(\ref{eqn:errcont}) that the error contour
will be aligned either parallel or perpendicular to the direction
of the neighbor, depending on whether $|k|$ is larger or smaller than
unity.  That is, $[{\cal R}(0)]^{-2} = 1 - k^2\exp[-(1+k)]$ and
$[{\cal R}(\pi/2)]^{-2} = 1 -\exp[-(1+k)]$.
Note that the contour will not be perfectly elliptical
but will generally approximate an ellipse.  In the present case
$a\sim 1.7\sigma$, so $k\sim 0.4$ and the contour is aligned perpendicular
to the direction of the neighbor.  However, since $\exp[-(1+k)]\sim 0.36$,
the overall deviation from circular symmetry is only about 14\% in this
case.

\section{Conclusions}
We measured the proper motion of the source star in the microlensing event
MOA-2011-BLG-262 
using new dipole-fitting method performed on the
difference image of two epochs separated by $\sim 8$ yrs.  
The result is $(-2.3, -0.9)\pm (2.8,2.6)\,\masyr$ 
in a (North, East) Galactic coordinate frame. 

The new method yields a measurement 
of proper motion of the star that is close to the photometric detection 
limit. In fact, the subject star was not even discovered in the initial 
photometry of the photometric reference image for the field, 
due to being faint and close to another star. The star's exact 
position and brightness was measured during the microlensing event.
Additionally, this new approach allows to marginalize over 
an unknown motion of the partially blended neighboring star. 

The importance of this measurement for our particular microlensing event
lies mainly in the fact that the obtained confidence regions
exclude the source being a high velocity star and show that it
follows the typical bulge kinematics.
It is compatible with both microlensing models -- and only 
increases {\it a priori} lensing probability of the very-high
$\mu_{\rel}$ microlensing model ($19.6\, \masyr$) by a factor
of 1.5 when compared to the high $\mu_{\rel}$ microlensing
model ($11.6\, \masyr$).
However, more importantly, it heavily disfavors lens-in-the-bulge scenario for the 
former and moderately disfavors it for the later.
The stellar lens with
the Jupiter-mass planetary companion located in the Galactic bulge is, 
{\it a priori}, much more likely explanation for the MOA-2011-BLG-262 microlensing event,
than the close-by Jupiter-mass planet with the Earth-mass moon in the Galactic disk.
By reducing the likelihood of the lens being in the bulge, our measurement
brings those two explanations on par.

For the microlensing model description and more detailed discussion 
of the solutions and impact our measurement has, see \citet{bennett14}.

\acknowledgements
J.S. acknowledge support of the Space Exploration 
Research Fund of The Ohio State University. 
The OGLE project has received funding from the 
European Research Council under the European
Community's Seventh Framework Programme (FP7/2007–2013)/ERC grant 
agreement No. 246678 to A.U.

%%%%%%%%%%%%%%%%%%%%%%%%%%%%%%%%%%%%%%%%%%%%%%%%%%%%%%%%%%%%%%%%%%%%%%%%%
%                           Bibliography
%%%%%%%%%%%%%%%%%%%%%%%%%%%%%%%%%%%%%%%%%%%%%%%%%%%%%%%%%%%%%%%%%%%%%%%%%

%                            Figures
%%%%%%%%%%%%%%%%%%%%%%%%%%%%%%%%%%%%%%%%%%%%%%%%%%%%%%%%%%%%%%%%%%%%%%%%%

\end{document}